\documentclass[conference]{IEEEtran}
\IEEEoverridecommandlockouts
\usepackage{cite}
\usepackage{amsmath,amssymb,amsfonts}
\usepackage{algorithmic}
\usepackage{graphicx}
\usepackage{textcomp}
\usepackage{xcolor}
\usepackage{import}
\usepackage{svg}
\usepackage{lipsum}
\usepackage{siunitx}
\usepackage{hyperref}
\usepackage{amsmath}
\usepackage{amssymb}
\usepackage{titlesec}
\usepackage{multicol}

\titlespacing\section{0pt}{5pt plus 2pt minus 2pt}{4pt plus 2pt minus 2pt}
\titlespacing\subsection{0pt}{4pt plus 2pt minus 2pt}{2pt plus 1pt minus 1pt}

\usepackage{adjustbox}
\usepackage{lengthconvert}
\Convertsetup{unit=cm}
\usepackage{bm}
\usepackage[acronym, shortcuts]{glossaries}

\makeglossaries
\newacronym{DBMC}{DBMC}{diffusion-based molecular communication}
\newacronym{TDMA}{TDMA}{time-division multiple access}
\newacronym{MDMA}{MDMA}{molecule-division multiple access}
\newacronym{NOMA}{NOMA}{non-orthogonal multiple access}
\newacronym{ADMA}{ADMA}{amplitude-division multiple access}
\newacronym{MA}{MA}{multiple access}
\newacronym{BEP}{BEP}{bit error probability}
\newacronym{TX}{TX}{transmitter}
\newacronym{RX}{RX}{receiver}
\newacronym{ISI}{ISI}{inter-symbol interference}
\newacronym{MAI}{MAI}{multiple-access interference}
\newacronym{SIC}{SIC}{successive interference cancellation}
\newacronym{OOK}{OOK}{on-off-keying}
\newacronym{MCS}{MCS}{Monte Carlo simulation}
\newacronym{SNR}{SNR}{signaling-molecule-to-noise ratio}
\newacronym{MC}{MC}{molecular communication}
\newacronym{IoBNT}{IoBNT}{internet of bio-nano-things}
\newacronym{BNM}{BNM}{bio-nano-machine}
\newacronym{DBMC-NOMA}{DBMC-NOMA}{NOMA for DBMC networks}
\newacronym{UCA}{UCA}{uniform concentration assumption}
\newacronym{SCRN}{SCRN}{stochastic chemical reaction network}
\newacronym{CRN}{CRN}{chemical reaction network}
\newacronym{ODE}{ODE}{ordinary differential equation}
\newacronym{SSA}{SSA}{stochastic simulation algorithm}
\newacronym{ChemSICal}{ChemSICal}{chemical reaction network for successive interference cancellation}
\newacronym{PDF}{PDF}{probability density function}
\newacronym{PMF}{PMF}{probability mass function}
\newacronym{RRC}{RRC}{reaction rate constant}

\glsdisablehyper

\def\BibTeX{{\rm B\kern-.05em{\sc i\kern-.025em b}\kern-.08em
    T\kern-.1667em\lower.7ex\hbox{E}\kern-.125emX}}

\begin{document}

\setlength{\abovedisplayskip}{5pt}
\setlength{\belowdisplayskip}{5pt}

\title{ChemSICal: Evaluating a Stochastic Chemical Reaction Network for Molecular Multiple Access\
\thanks{The authors acknowledge the financial support by the Federal Ministry of Education and Research of Germany in the program of “Souverän. Digital. Vernetzt.”. Joint project 6G-life, project identification number: 16KISK002.}
\vspace{-0.3cm}
}

\author{\IEEEauthorblockN{Alexander Wietfeld, Marina Wendrich, Sebastian Schmidt, Wolfgang Kellerer}
\IEEEauthorblockA{\textit{Chair of Communication Networks} \\
\textit{Technical University of Munich}, Germany\\
\{alexander.wietfeld, marina.wendrich, sebastian.a.schmidt, wolfgang.kellerer\}@tum.de}
}

\maketitle

\begin{abstract}

Proposals for \acl{MC} networks as part of a future \acl{IoBNT} have become more intricate and the question of practical implementation is gaining more importance. One option is to apply detailed chemical modeling to capture more realistic effects of computing processes in biological systems. In this paper, we present ChemSICal, a detailed model for implementing the \acf{SIC} algorithm for molecular \acl{MA} in \acl{DBMC} networks as a \acf{CRN}. We describe the structure of the model as a number of smaller reaction blocks, their speed controlled by \acfp{RRC}. Deterministic and stochastic methods are utilized to first iteratively improve the choice of \acsp{RRC} and subsequently investigate the performance of the model in terms of an error probability. We analyze the model's sensitivity to parameter changes and find that the analytically optimal values for the non-chemical model do not necessarily translate to the chemical domain. This necessitates careful optimization, especially of the \acsp{RRC}, which are crucial for the successful operation of the ChemSICal system. 

\end{abstract}

\begin{IEEEkeywords}
molecular communication, chemical reaction networks, NOMA, successive interference cancellation
\end{IEEEkeywords}

\section{Introduction}\label{sec:introduction}

\Ac{DBMC} is a communication paradigm that utilizes molecules as information carriers. As a biocompatible, energy-efficient and natively nano-scale solution, \ac{DBMC} could facilitate information exchange in a future \ac{IoBNT}, where biological environments like the human body are accessible to networking. The realization of the \ac{IoBNT} could enable revolutionary medical use case such as targeted drug delivery or advanced monitoring and diagnosis~\cite{akyildizInternetBioNanoThings2015, HofmannMolecularPerspective2024}.
Nodes within these \ac{DBMC} networks are expected to be natural or synthetic structures on the micro- and nanoscale. The field of bioengineering has seen significant progress in recent years and engineered cells or bacteria can function as devices with specific but limited capabilities~\cite{akyildizInternetBioNanoThings2015}. To realize the envisioned complex applications, the establishment of \ac{DBMC} networks for communication and cooperation is necessary. 

\subsection{State of the Art}
As one of the first steps towards more communication participants, \ac{MA} has been an important topic of investigation for \ac{DBMC} research. Proposals include \ac{TDMA}~\cite{shitiriTDMABasedDataGathering2021}, where \acp{TX} are assigned individual time slots and have to wait for their turn to transmit, or \ac{MDMA}~\cite{chenResourceAllocationMultiuser2021}, where each \ac{TX} uses a different molecule type to communicate and the \ac{RX} can, therefore, differentiate them chemically. To combine the simultaneous transmissions from \ac{MDMA} and the use of only a single molecule type from \ac{TDMA}, \ac{NOMA} has been proposed as a possible solution~\cite{wietfeldDBMCNOMAEvaluatingNOMA2024}.
To realize \ac{NOMA}, the implementation of \ac{SIC} at the \ac{RX} is necessary to differentiate the \acp{TX} based on the number of received molecules.
\Ac{DBMC-NOMA} has been investigated from an analytical point of view~\cite{wietfeldDBMCNOMAEvaluatingNOMA2024}. However, the implications of implementing such an algorithm in a biological environment within tiny low-capability devices has not been addressed.
\begin{figure}[t]
    \centering
    \includegraphics[width=0.55\linewidth]{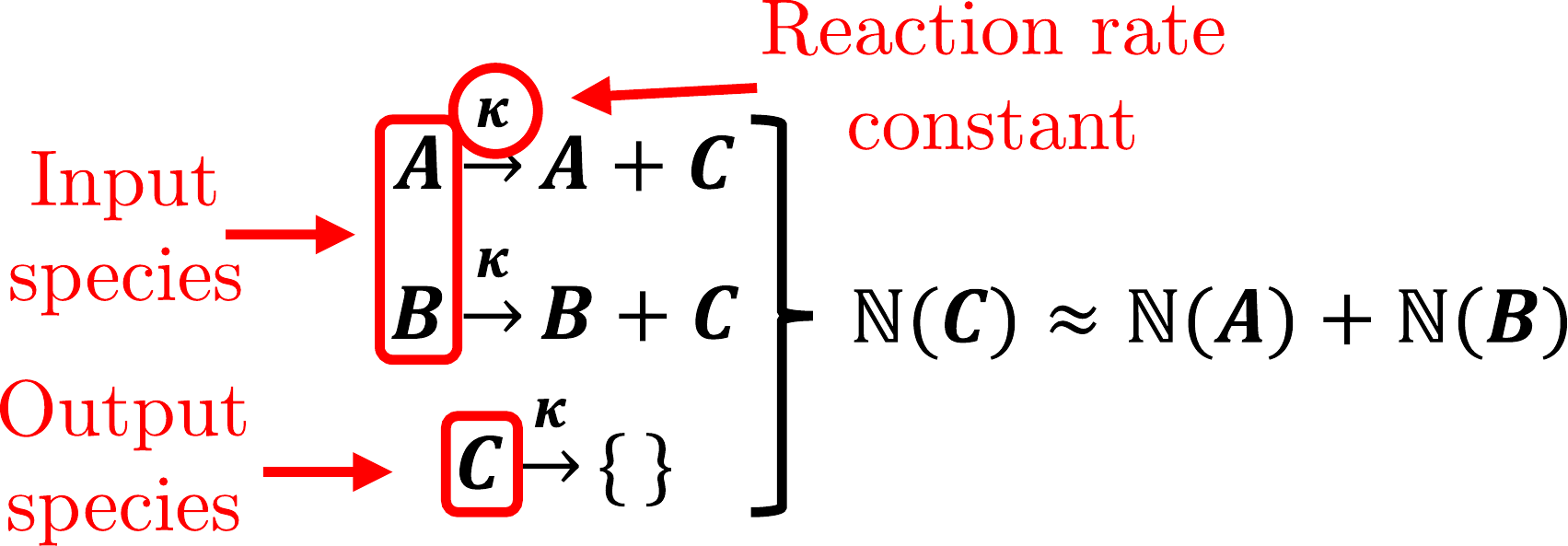}
    \vspace{-0.3cm}
    \caption{Example of a simple chemical reaction network computing the sum of two inputs $A$ and $B$ as a third chemical species $C$.}
    \label{fig:crn_example}
    \vspace{-0.5cm}
\end{figure}
In contrast to classical communication systems, \ac{DBMC} networks will not have access to general-purpose computing resources. As a result, the notion of \acp{CRN} has gained significant research attention. 
Vasić et al. discuss the capabilities of \acp{CRN} and describe several examples, which are capable of executing simple functionalities like comparison or addition~\cite{vasic_crn_2020}. An example of a simple \ac{CRN} can be seen in Figure~\ref{fig:crn_example}. Each reaction is described by reactants or input species, products or output species, and a \ac{RRC} that determines the speed with which the reaction occurs. In this case, the species $A$ and $B$ form the inputs, and their concentrations, denoted as $\mathbb{N}(A), \mathbb{N}(B)$ together form the concentration of the output species $C$, as $\mathbb{N}(C) = \mathbb{N}(A) + \mathbb{N}(B)$, after the system has reached a steady-state, thereby implementing addition.
Other researchers have specifically targeted \ac{DBMC} networks. For example, Bi et al. investigated and implemented several simple functionalities such as logic gates and amplification in the form of \acp{CRN} for use in a molecular communication node.
Since chemical reactions are governed by stochastic processes, deterministic models do not capture their behavior fully, and it is important to also utilize \ac{SCRN} models~\cite{heinleinClosingImplementationGap2024b}. In~\cite{heinleinClosingImplementationGap2024b}, Heinlein et al. model a \ac{RX} with multiple functionalities such as detection and synchronization solely using \acp{SCRN}. Angerbauer et al. have used chemical reactions to model simple neural network architectures that could be used in future \ac{DBMC} communication nodes~\cite{angerbauerMolecularNanoNeural2024}.
However, \acp{CRN} have not previously been used in the context of \ac{DBMC} networks to implement multi-node capabilities such as \ac{MA}, which will be crucial in the \ac{IoBNT}.

    \subsection{Contributions}
In this work, we present the following contributions.
    \begin{itemize}
        \item We propose ChemSICal, a model for chemical implementation of SIC for \ac{MA} in a simple DBMC network, expanding our initial proposal in~\cite{wietfeldChemicalReactionNetwork2024a}.
        \item We use deterministic and stochastic methods to optimize the network and evaluate the impact ChemSICal has on the algorithm performance.
        \item We conduct a sensitivity analysis of selected parameters to show that the \acp{RRC} values are critical for the system's performance and that analytically optimal parameters might not be optimal in the chemical model. It will be necessary to adapt the algorithm to the chemical implementation, even at the cost of deviating from the analytically optimal parameters.
    \end{itemize}

\section{System Model}\label{sec:system_model}

\begin{figure}
    \centering
    \includegraphics[width=0.7\linewidth]{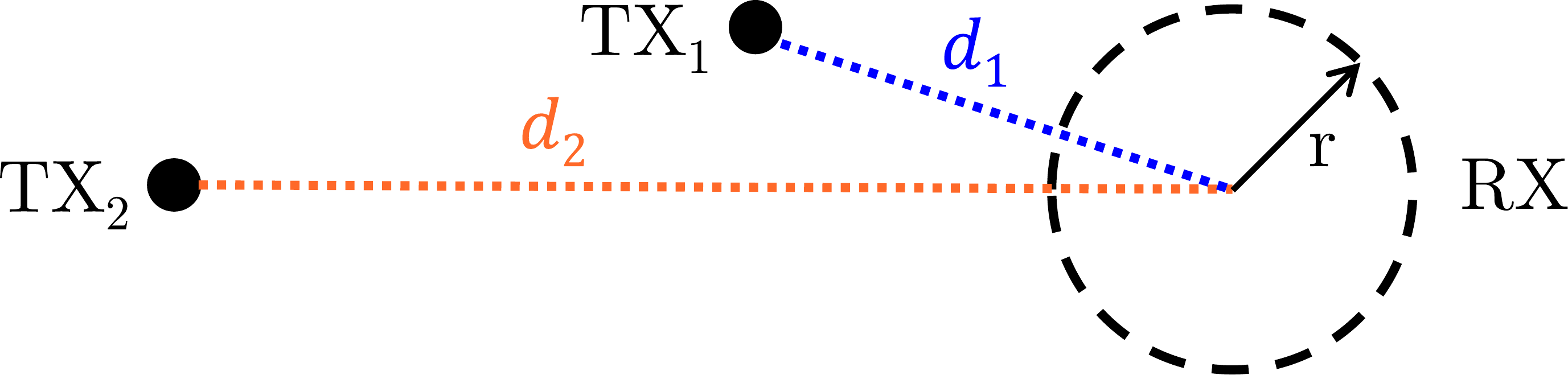}
    \vspace{-0.3cm}
    \caption{Simple \ac{DBMC} network with 2 point \acp{TX} and a passive spherical \ac{RX}}
    \label{fig:scenario}
    \vspace{-0.5cm}
\end{figure}
In the following, a communication scenario is presented forming the underlying framework in which the ChemSICal model will be tested. Additionally, the concepts of \ac{DBMC-NOMA} and the \ac{SIC} algorithm will be explained.

\subsection{Communication Scenario}\label{subsec:scenario}

To put the proposed \ac{SIC} model in context, we look at a simplified communication scenario, in which we can employ ChemSICal on the \ac{RX} side. In the following, we will describe the underlying assumptions regarding the network and communication system.
In Figure \ref{fig:scenario}, a \ac{DBMC} network  is depicted with 2 \acp{TX} at distances $d_1 \leq d_2$ from a single \ac{RX}. The \ac{RX} is modeled as a passive spherical observer of the molecules within its receiving volume $V_\mathrm{RX} = \frac{4}{3}\pi r^3$, where $r$ is the \ac{RX} radius. The \acp{TX} are assumed to be 1D points capable of instantaneously emitting pulses of $N_\mathrm{TX}$ molecules.
Given a free diffusion channel with diffusion coefficient $D$, the average received number of molecules over time at the \ac{RX}, $\lambda_i(t)$, following a pulse from TX$_i$ at time $t=0$ is~\cite{jamaliChannelModelingDiffusive2019}
 \begin{equation}
     \lambda_i(t) = \frac{N_{\mathrm{TX}}V_\mathrm{RX}}{\left(4\pi Dt\right)^\frac{3}{2}}\exp \left(-\frac{d_i^2}{4Dt}\right)\ \mathrm{for}\ t\geq0.  
\end{equation}

Furthermore, we assume that the \acp{TX} use \ac{OOK} to transmit binary information symbols $s_i\in\{0,1\}$ by emitting none or $N_\mathrm{TX}$ molecules, respectively. The system is fully synchronized and for the sake of simplicity, we neglect the \ac{ISI} on the basis that the symbol period between subsequent pulses must be sufficiently large.
As a result, we will focus on a single symbol period, in which the \ac{RX} takes a single sample $n_\mathrm{s}$ at the peak time $t_\mathrm{p}$ of the received number of molecules.
In this work, the \ac{DBMC-NOMA} scheme with \ac{SIC}, as defined in~\cite{wietfeldDBMCNOMAEvaluatingNOMA2024} and described in Section~\ref{subsec:sic}, is implemented. Therefore, both \acp{TX} transmit at the same time, denoted as $t=0$, using the same molecule type. As diffusion is a stochastic process, the instantaneous received number of molecules is commonly modeled as a Poisson-distributed random variable~\cite{torresgomezAgeInformationMolecular2022}. The \ac{RX} will observe the sum of the two independent Poisson variables corresponding to the received signal from each \ac{TX}. The sum of two independent Poisson variables is a Poisson variable and the sample at the \ac{RX} is therefore, $n_\mathrm{s}\sim \mathcal{P}\left(s_1\lambda_1(t_p) + s_2\lambda_2(t_p)\right)$.
The probability distribution of $n_\mathrm{s}$ can be further specified, given that the four different symbol combinations in $\{[s_1 s_2], s_1,s_2\in\{0,1\}\}$ are assumed to be equiprobable. Then, the \ac{PMF} for the sample at the \ac{RX} is given by
\begin{multline}\label{eq:input_pdf}
     p_{n_\mathrm{s}}[n] = \frac{1}{4}\cdot\left(\mathcal{P}_\mathrm{PMF}(n;0) + \mathcal{P}_\mathrm{PMF}(n;\lambda_2)\right. \\
     \left. + \mathcal{P}_\mathrm{PMF}(n;\lambda_1) + \mathcal{P}_\mathrm{PMF}(n;\lambda_1+\lambda_2)\right),
\end{multline}
where $\mathcal{P}_\mathrm{PMF}(n;\lambda) = \lambda^n\frac{e^{-\lambda}}{n!}$ is the \ac{PMF} of the Poisson distribution at $n$.

\subsection{NOMA and Successive Interference Cancellation}\label{subsec:sic}

\begin{figure}
    \centering
    \includegraphics[width=0.45\linewidth]{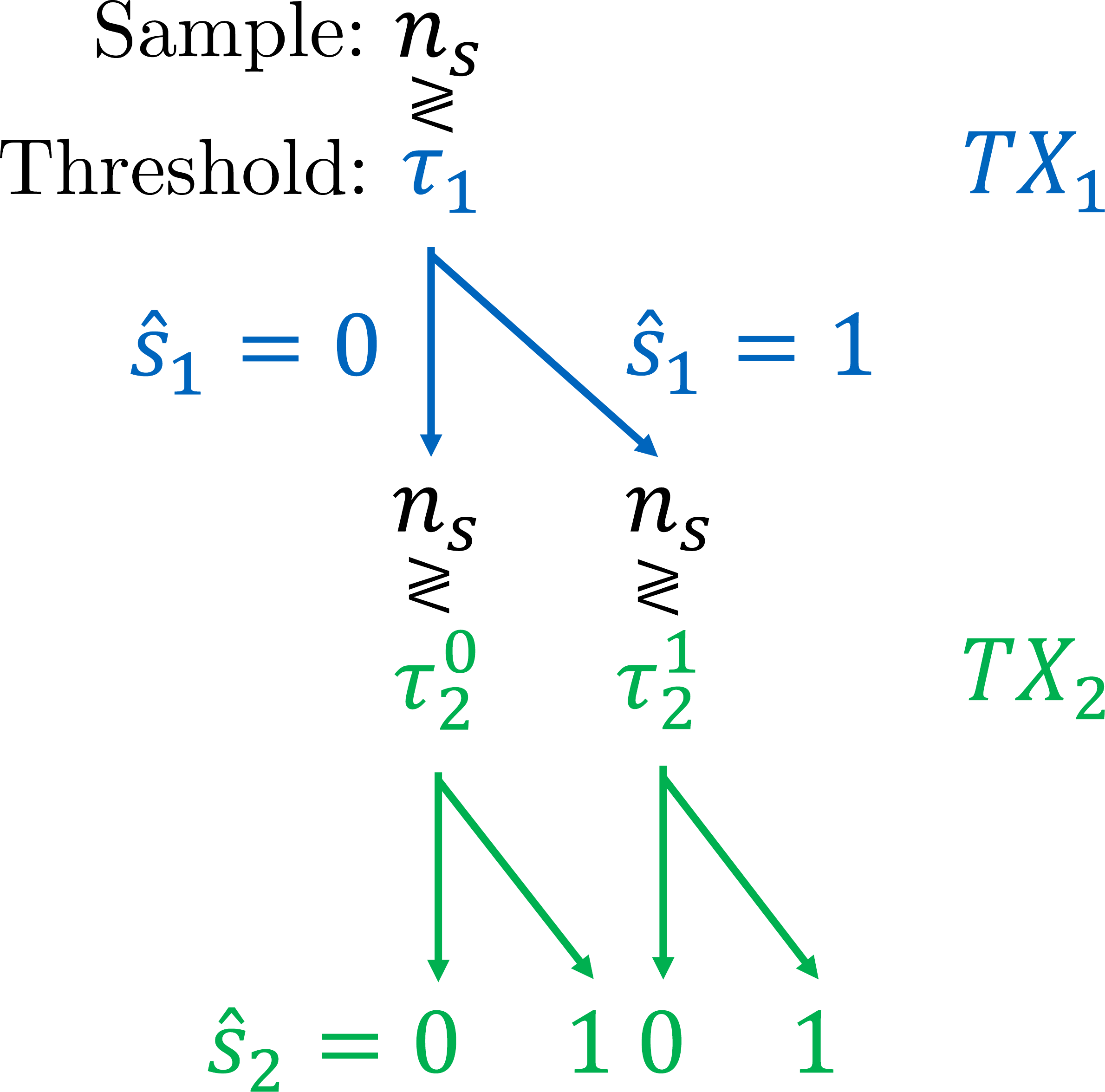}
    \vspace{-0.3cm}
    \caption{Simplified \ac{SIC} algorithm for \ac{DBMC}~\cite{wietfeld_error_2024}. Example for 2 \acp{TX}.}
    \label{fig:sic_diagram}
    \vspace{-0.6cm}
\end{figure}

The theoretical basis for symbol differentiation at the \ac{RX} is the \ac{SIC} algorithm as presented in~\cite{wietfeld_error_2024}. An adaptive threshold detection system is employed resulting in a binary tree decision structure depicted in Figure~\ref{fig:sic_diagram}. It is based on the assumption that the received numbers of molecules from the \acp{TX} differ in magnitude. This can be facilitated through different mechanisms. One option is to optimize the numbers of emitted molecules, for example, using a pilot-symbol-based algorithm~\cite{wietfeld_error_2024}. In order to focus primarily on the \ac{CRN} implementation, consideration of the optimization algorithm in the chemical domain is left for future work. In this investigation we will assume that due to the random diffusive nature of \ac{DBMC} systems, it is very unlikely for two \acp{TX} to be at the same distance from the \ac{RX} at a particular time. Therefore, the average received numbers of molecules obey the relation $\lambda_1 > \lambda_2$, due to the assumption that $d_1 > d_2$~\cite{wietfeld_non-orthogonal_2023}.
The \ac{SIC} algorithm then works as follows.
The acquired sample $n_\mathrm{s}$ is compared in multiple steps. In the first step, the symbol for TX$_1$, $\hat{s}_1$, is detected and in the second step, the threshold depends on the first step, yielding the symbol $\hat{s}_2$ for TX$_2$, as shown in the following: 
\begin{multicols}{2}
    \noindent
    \begin{equation}
        \hat{s}_1 = \begin{cases}
            1   &n_\mathrm{s}\geq \tau_1\\
            0   &\mathrm{otherwise}.
        \end{cases}
    \end{equation}
    \begin{equation}
        \hat{s}_2 = \begin{cases}
            1   &n_\mathrm{s}\geq \tau_1^{\hat{s}_1}\\
            0   &\mathrm{otherwise}.
        \end{cases}
    \end{equation}
\end{multicols}

This structure is theoretically extendable to arbitrary numbers of \acp{TX}~\cite{wietfeld_error_2024} but we focus on the case of 2 \acp{TX} for this work.

\section{ChemSICal Design}

The analysis of a network using \ac{DBMC-NOMA} has been conducted in~\cite{wietfeld_non-orthogonal_2023, wietfeldDBMCNOMAEvaluatingNOMA2024}, yielding optimal values for the detection thresholds given a certain distance constellation, but without considering the chemical implementation of the \ac{SIC} algorithm. Therefore, our aim is to frame the algorithm as a chemical computing problem and to propose ChemSICal, a chemical reaction network for \ac{SIC}.
As detailed in Section~\ref{sec:introduction}, a \ac{CRN} is a set of coupled chemical reactions consisting of different species of molecules, reactant-product relations, and \acp{RRC}. Firstly, we will define the desired input and output species.
To create a link between the underlying \ac{DBMC-NOMA} system and the \ac{CRN} on the \ac{RX} side, the signaling molecule type is defined to be of the input species $Y_\mathrm{on}$. Consequently, the number of molecules $n_\mathrm{s}$ sampled by the \ac{RX} as detailed in Section~\ref{subsec:scenario} is used as the initial concentration of the $Y_\mathrm{on}$ molecules for the \ac{CRN}. We denote the species by its name, e.g., $Y_\mathrm{on}$, and the corresponding number of molecules with $\mathbb{N}(Y_\mathrm{on})$.
After a number of chemical reactions occur, as will be described in the next section, we end up with output species $D_i^0$, $D_i^1$ with $i\in\{1,2\}$ for both \acp{TX}. The symbols detected by the \ac{CRN}, $\hat{s}_i^C$, are determined via the number of output species molecules as follows:
    \begin{equation}
        \hat{s}_i^C = \begin{cases}
            1 &\mathbb{N}(D_i^1) \geq \mathbb{N}(D_i^0)\\
            0 &\mathbb{N}(D_i^0) > \mathbb{N}(D_i^1).
        \end{cases}
    \end{equation}
    \vspace{-0.3cm}

\begin{figure*}[t]
    \centering
    \includegraphics[width=\linewidth]{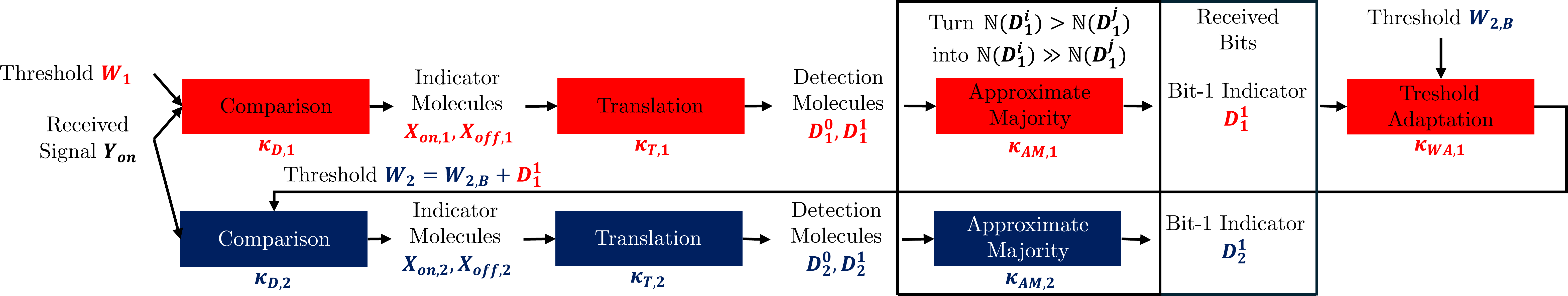}
    \vspace{-0.3cm}
    \caption{Proposed ChemSICal structure: A \ac{CRN} implementing a two-stage \acs{SIC} algorithm on the \ac{RX} side. Input: $Y_\mathrm{on}$; Outputs: $D_i^0$, $D_i^1$, $i\in\{1,2\}$}
    \label{fig:blockdiagram}
    \vspace{-0.5cm}
\end{figure*}

\subsection{CRN Block Diagram}~\label{subsec:crn_structure}

Figure~\ref{fig:blockdiagram} depicts the logical structure of ChemSICal. In order to keep the overview of the proposal as simple as possible, as well as collect related reactions in logically coherent units, we utilize reaction blocks, capable of executing simpler operations such as comparisons or additions, each consisting of up to four individual chemical reactions. To implement ChemSICal, we need the following reaction blocks, as described in~\cite{vasic_crn_2020}:
    \begin{itemize}
        \item \textit{Comparison:} Compares an input species ($Y_\mathrm{on}$) to a threshold ($W_i$) and generates two indicator molecule species ($X_{\mathrm{on},i}$, $X_{\mathrm{off},i}$), the ratio of which indicates the ratio between input and threshold.
        \item \textit{Translation:} From one or multiple inputs ($X_{\mathrm{on},i}$, $X_{\mathrm{off},i}$), generates detection molecule species ($D_i^0$, $D_i^1$) of the exact same number. This is used to decouple subsequent reactions, i.e., limit interference of the previous reaction on the reactants of the following ones.
        \item \textit{Approximate Majority:} Takes two inputs ($D_i^0$, $D_i^1$) and greatly amplifies any differences in numbers between them, i.e., turns $\mathbb{N}(D_i^j) > \mathbb{N}(D_i^k)$ into outputs $\mathbb{N}(D_i^j) \gg \mathbb{N}(D_i^k)$. This results in an approximate binary decision, where many more of one species, and almost none of the other are left.
        \item \textit{Threshold Adaptation:} Essentially corresponds to an addition \ac{CRN}, as shown in Figure~\ref{fig:crn_example}, where the base threshold species, $W_{2,\mathrm{B}}$, and the resulting detection species $D_1^1$ are added to form the threshold species $W_2$. This links the first stage for detecting the bit from TX$_1$ with the second stage for the bit from TX$_2$. The numbers from each species must be tuned such that the following holds: $\mathbb{N}(W_{2,\mathrm{B}}) = \tau_2^0$ and $\mathbb{N}(X_{\mathrm{on},1}) + \mathbb{N}(X_{\mathrm{off},1}) + \mathbb{N}(W_{2,\mathrm{B}}) = \tau_2^1$. The latter must hold, since $\mathbb{N}(D_1^1) \approx \mathbb{N}(X_{\mathrm{on},1}) + \mathbb{N}(X_{\mathrm{off},1})$ after the first-stage \textit{Approximate Majority} block in case of $\hat{s}_1^C = 1$.
    \end{itemize}
Each reaction block is associated with a particular \ac{RRC}, as denoted under the blocks in Figure~\ref{fig:blockdiagram}. Every reaction in each block uses this \ac{RRC}.
The exact reactions with intermediary molecule species can be found in~\cite{wietfeldChemicalReactionNetwork2024a}, and are adapted from~\cite{vasic_crn_2020}.
Importantly, the sequential nature implied by the arrows in Figure~\ref{fig:blockdiagram} is purely logical and to make the diagram more easily understandable. In fact, the reaction are all started at the same time and run in parallel to each other. Therefore, no additional external control over the \ac{CRN} is required.

\subsection{Solvers and Evaluation Metrics}\label{subsec:metrics}

In the following, we will define the metrics used for the subsequent evaluation of the ChemSICal model. They are based on the detection outcome $\hat{s}_i^C$, as defined above.
The correctness of the result depends on the input value $Y_\mathrm{on}$, i.e. the underlying received signal $n_\mathrm{s}$, which dictates the expected outcome according to the non-chemical \ac{SIC} algorithm in Figure~\ref{fig:sic_diagram}.
Additionally, the appropriate metric depends on the type of solver that is applied to the \ac{CRN}. We will use both a deterministic and a stochastic solver for the evaluation.
The deterministic solver converts the entire \ac{CRN} into a set of \acp{ODE} and finds the solution as a continuous time-varying function of each species' concentration~\cite{abelGillesPyPythonPackage2016}. The deterministic approaches cannot capture the probabilistic dynamics of \acp{CRN}, especially in cases where the molecule concentration is low~\cite{abelGillesPyPythonPackage2016}. Therefore, we also apply a \ac{SSA} solver, which uses a time-continuous value-discrete Markov process and the well-known Gillespie's algorithm~\cite{gillespie_exact_1977}. The latter generates a statistically accurate sample trajectory of the system~\cite{abelGillesPyPythonPackage2016}. Subsequently, a Monte Carlo simulation using $N_\mathrm{traj}$ trajectories is conducted to gain more insight into the statistical properties of the system. The following metrics will be used:
    \begin{enumerate}
        \item \textit{Correct Detection}: We define the binary indicator for a correct outcome given a particular input $\mathbb{N}(Y_\mathrm{on})$ as equal to 1 if the outcome $[\hat{s}_1^C \hat{s}_2^C]$ is equal to $[\hat{s}_1 \hat{s}_2]$, and 0 if it is incorrect. For the \ac{ODE} solver, this is denoted as $c[\mathbb{N}(Y_\mathrm{on})]\in\{0,1\}$, or for a single trajectory $j$ generated by the \ac{SSA} solver, as $c_j[\mathbb{N}(Y_\mathrm{on})]\in\{0,1\}$.
        \item \textit{Probability of Correct Detection}: This corresponds to the percentage of trajectories that result in a correct result, given a particular input $\mathbb{N}(Y_\mathrm{on})$. It is denoted and calculated as $P_\mathrm{d}[\mathbb{N}(Y_\mathrm{on})] = \frac{1}{N_\mathrm{traj}}\sum_{j=1}^{N_\mathrm{traj}} c_j[\mathbb{N}(Y_\mathrm{on})]$.
        \item \textit{Input-Weighted Probability of Error}: The errors must be weighted according to the \ac{PMF} given in Eq. (\ref{eq:input_pdf}), such that errors for unlikely inputs are less influential compared to errors for inputs that occur often. We denote the input PMF as $p_{n_\mathrm{s}}[\mathbb{N}(Y_\mathrm{on})]$. The weighted error probability can be calculated as $P_\mathrm{e} = \sum_{\mathbb{N}(Y_\mathrm{on})} p_\mathrm{n_\mathrm{s}}[\mathbb{N}(Y_\mathrm{on})]\cdot P_\mathrm{d}[\mathbb{N}(Y_\mathrm{on})]$.
    \end{enumerate}

\section{Evaluation and Optimization}\label{sec:stochastic_eval}

In the following, we will take a closer look at the performance of the ChemSICal model using the \ac{ODE} and \ac{SSA} solver. Recall that a non-chemical version of \ac{DBMC-NOMA} and a simple \ac{DBMC} network was described in Section~\ref{sec:system_model}. For the following evaluation, the parameters in Table~\ref{tab:parameters} define the communication scenario.
The detection thresholds are assumed to be chosen optimally using exhaustive search for the given scenario~\cite{wietfeldDBMCNOMAEvaluatingNOMA2024}. 
Other options such as heuristic optimization using pilot symbols~\cite{wietfeld_error_2024} have also been proposed.
Using results from~\cite{wietfeldDBMCNOMAEvaluatingNOMA2024}, the \ac{BEP} of the non-chemical network is approximately $P_\mathrm{e,ideal} \approx 10^{-6}$.
Additionally, the corresponding chemical parameters, initial concentrations and \acp{RRC}, are listed in Table~\ref{tab:initia_values}. Unless otherwise stated, the parameters marked as \textit{baseline} will be used. These correspond to the parameters that are translated from the optimal non-chemical parameters according to the relationships for detection thresholds derived in Section~\ref{subsec:crn_structure}.

\begin{table}[t]
\caption{Communication System Parameters}
\vspace{-0.2cm}
\label{tab:parameters}
\centering
\resizebox{0.9\columnwidth}{!}{%
\begin{tabular}{lll}
\hline
\textbf{Parameter}    & \textbf{Symbol}                            & \textbf{Values}                          \\ \hline
TX distances          & $\{d_1, d_2\}$                          & $\{10,12\}\,\qty{}{\micro\meter}$                \\
RX radius             & $r$                                        & $\qty{1}{\micro\meter}$                 \\
Diffusion coefficient & $D$                                        & $\qty{e-9}{\meter\squared\per\second}$ \\
Molecules per pulse                                & $N_\mathrm{TX}$ & $\qty{e6}{\mathrm{molecules}}$                                  \\ 
Detection thresholds                      & $\{\tau_1, \tau_2^0, \tau_2^1\}$  & \{231, 78, 386\}\,$\qty{}{\mathrm{molecules}}$           \\
\hline
\end{tabular}%
}
\vspace{-0.2cm}
\end{table}

\begin{table}[t]
\centering
\caption{\ac{CRN} Initial Values and Parameters}
\vspace{-0.2cm}
\label{tab:initia_values}
\resizebox{0.95\columnwidth}{!}{%
\begin{tabular}{lll}
\hline
Parameter          & Symbol          & Initial Value (\underline{Baseline}) \\ \hline
Input Species      & $Y_{on}$        & $0\leq \mathbb{N}(Y_\mathrm{on}) <600$                  \\
Result Species     & $X_{on,1}$      & \{\underline{154}, 155, 156, 157\}                   \\
                   & $X_{off,1}$     & \{\underline{154}, 155, 156, 157\}                   \\
                   & $X_{on,2}$      & 83                    \\
                   & $X_{off,2}$     & 84                    \\
Detection Species  & $\{D_1^0, D_1^1, D_2^0, D_2^1\}$       & \{0,0,0,0\}                     \\
Threshold Species  & $W_{1}$         & 231                   \\
                   & $W_{2}$         & \{75, 76, 77, \underline{78}\}                   \\
                   & $W_{2,B}$          & \{75, 76, 77, \underline{78}\}                    \\
Helper Species     & $\{B_{1}, B_2\}$         & \{0,0\}                     \\
\acp{RRC}     & $\kappa_\mathrm{AM,2}$ & \{0.8, \underline{1}, 1.2, 1.4, 1.6, 3\}$\cdot 10^{-3}$\\
                   & Other $\kappa_i$       & See set 5 in Table \ref{tab:ODE_reaction_rates}\\
                   
\hline

\end{tabular}%
}
\vspace{-0.4cm}
\end{table}

\begin{figure}
    \centering
    \includegraphics[width=0.85\linewidth]{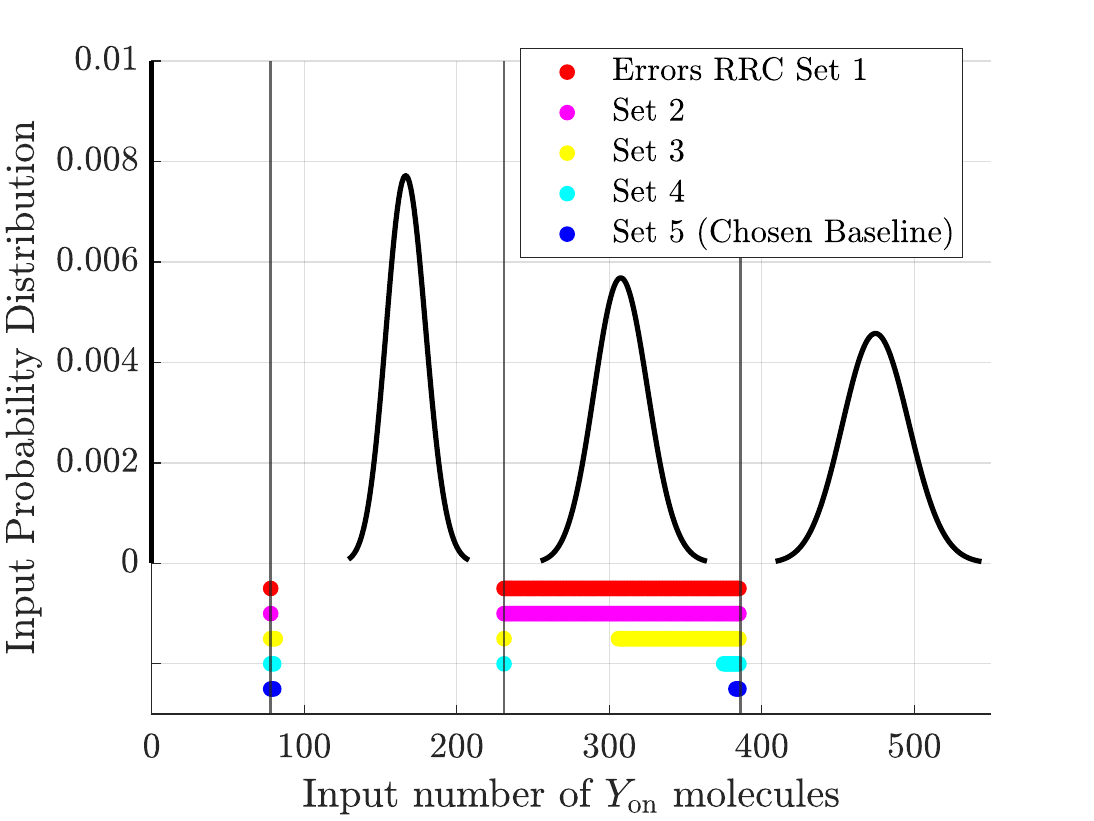}
    \vspace{-0.3cm}
    \caption{Comparison of error distributions obtained from the \ac{ODE} solver for different \ac{RRC} sets (see Table~\ref{tab:ODE_reaction_rates}). \textit{Set 5} is chosen as baseline for further evaluation. The input probability distribution (see Eq. (\ref{eq:input_pdf}) is also shown.}
    \label{fig:rate_optimization}
    \vspace{-0.5cm}
\end{figure}

\begin{figure}[t]
    \centering
    \includegraphics[height=4cm]{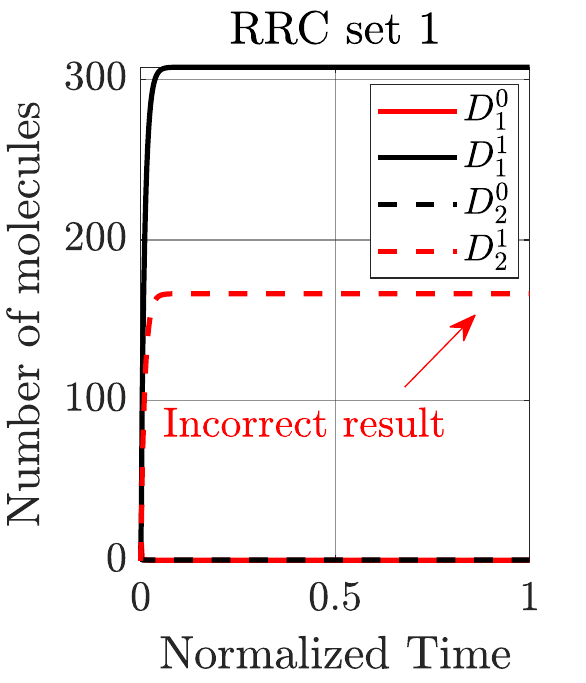}
    \includegraphics[height=4cm]{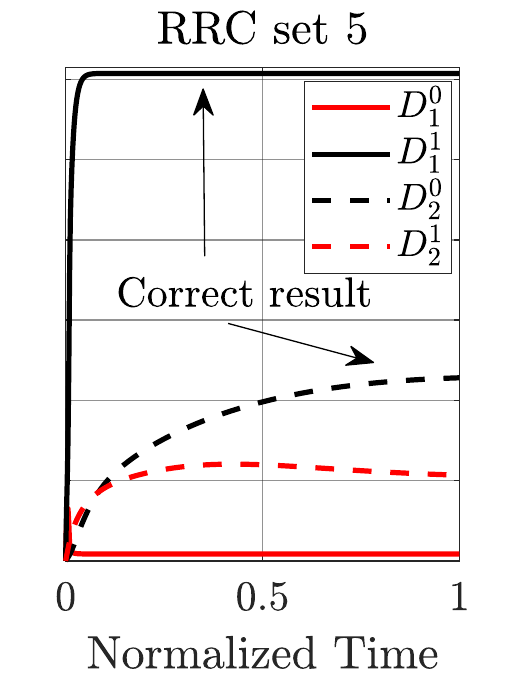}
    \vspace{-0.3cm}
    \caption{Comparison of two different sets of \acp{RRC} from Table~\ref{tab:ODE_reaction_rates} with respect to the detection species trajectories. The input is $\mathbb{N}(Y_\mathrm{on}) = 300$ and the desired result is for $D_1^1$ and $D_2^0$ (in black) to be in the majority.}
    \label{fig:reaction_rates}
    \vspace{-0.3cm}
\end{figure}

\begin{table}[t]
\caption{Reaction constant sets utilized in Figure \ref{fig:rate_optimization}.}
\vspace{-0.2cm}
\label{tab:ODE_reaction_rates}
\resizebox{\columnwidth}{!}{%
\begin{tabular}{lllllllll}
\hline
Set & $\kappa_{D,1}$ & $\kappa_{T,1}$ & $\kappa_{AM,1}$ & $\kappa_{WA,1}$ & $\kappa_{D,2}$ & $\kappa_{T,2}$ & $\kappa_{AM,2}$ & ODE Errors \\ \hline
1          & 1.0            & 1.0            & 1.0             & 1.0             & 1.0            & 1.0            & 1.0             & 0.25       \\
2          & 1.0            & 1.0            & 1.0             & 1.0             & 0.1            & 0.1            & 0.1             & 0.25        \\
3          & 1.0            & 1.0            & 0.1             & 1.0             & 0.1            & 0.1            & 0.01            & 0.1382       \\
4          & 1.0            & 1.0            & 0.1             & 1.0             & 0.1            & 0.01           & 0.001           & $3.03\cdot 10^{-3}$        \\
5          & 1.0            & 1.0            & 0.1             & 1.0             & 0.1            & 0.1            & 0.001           & $3.98\cdot 10^{-4}$       \\ \hline
\end{tabular}%
}
\vspace{-0.4cm}
\end{table}

\subsection{Reaction Rate Constants}\label{subsec:rate_optimization}

The first crucial evaluation step is the choice of \acp{RRC} for each reaction block.
As the design space here is very large and the resources needed for systematic optimization are quite extensive, in this work we focus on a first preliminary choice of the \acp{RRC} based on iteratively improving the number of correct detections $c[\mathbb{N}(Y_\mathrm{on})]$ using the \ac{ODE} solver. 
Figure~\ref{fig:rate_optimization} depicts the observed errors over the different input values for five different sets of \acp{RRC}. The associated values can be seen in Table~\ref{tab:ODE_reaction_rates}.

In all our evaluations, we use a normalized time axis as well as normalized \acp{RRC} relative to 1, as all \acp{RRC} could be multiplied by the same scalar value and all the results would simply be scaled on the time axis.
The starting point, \textit{set 1}, is setting all \acp{RRC} $\kappa_i = 1$. Here, we see a significant number of errors over a large section of the input space including very likely values, where $[\hat{s}_1 \hat{s}_2] = [1 0]$. This is largely due to the fact that the reactions for TX$_2$ will start computing the binary decision before the threshold adaptation reaction has correctly increased $\mathbb{N}(W_2)$. The input is compared to the wrong threshold and the \ac{CRN} returns $\hat{s}_2^C=1$ incorrectly. Once the approximate majority block has reached a steady-state it is very hard to reverse it. For higher values of $\mathbb{N}(Y_\mathrm{on})$ the errors seemingly disappear, but the underlying issue still remains. The result $\hat{s}_2^C=1$ is now merely accidentally correct, while the input is still being compared to the wrong value of $\mathbb{N}(W_2)$.
Generally, the goal is compute the values within the \ac{CRN} in the correct order, making reactions happen slower that work with results from previous reaction blocks as inputs. Looking at \textit{sets 2} and~\textit{3}, we can see that the issue is not solved by simply slowing down all the reactions for TX$_2$ (\textit{set 2}), and neither by slowing down the approximate majority blocks by an order of magnitude (\textit{set 3}), although the results for \textit{set 3} suggest that we are moving in the correct direction, with some errors disappearing. Lastly, \textit{set 4} and \textit{5} suggest that the approximate majority reaction for TX$_2$ should be slowed down even further, while the other reactions associated with TX$_2$ remain at an only slightly slower speed. A direct comparison of the detection species' trajectories calculated by the \ac{ODE} solver between \textit{set 1} and \textit{set 5} is shown in Figure~\ref{fig:reaction_rates} for $\mathbb{N}(Y_\mathrm{on}) = 300$. The results for \textit{set 1} on the left show that $D_2^1$ incorrectly dominates because the second comparison and approximate majority block reacts too quickly to the non-adapted incorrect threshold. \textit{Set 5} on the right side corrects the initial rise of $D_2^1$ and yields the correct result. As a consequence, we will use \textit{set 5} as the baseline choice of \acp{RRC} going forward.

\begin{figure}[t]
    \centering
    \includegraphics[width=0.9\linewidth]{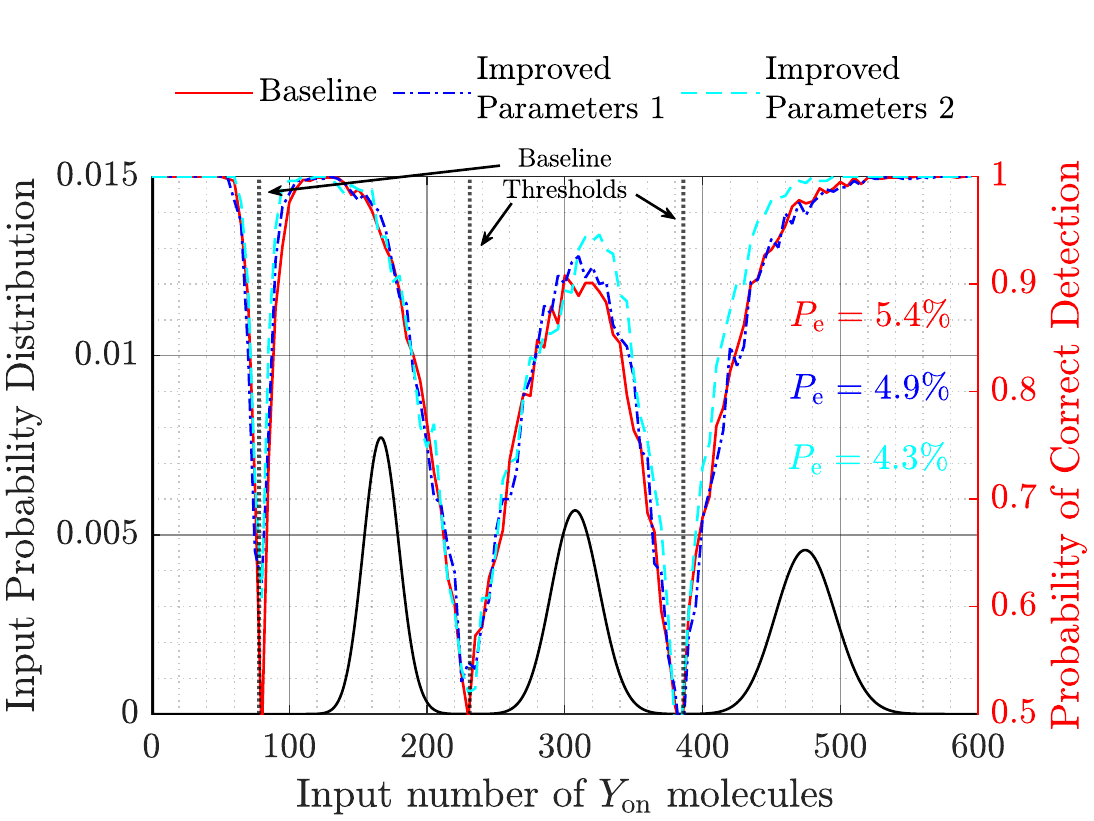}
    \vspace{-0.3cm}
    \caption{Plot showing the input \ac{PMF} and the probability of correct detection $P_d$ over the input values $\mathbb{N}(Y_\mathrm{on})$ for $N_\mathrm{traj}=500$ using the \ac{SSA} solver. For \textit{Improved Parameters 1}, the chemical species for the detection thresholds have been adjusted to $\{\tau2^0, \tau_2^1\} = \{76, 388\}$. For \textit{Improved Parameters 2}, the thresholds as well as the \ac{RRC} $\kappa_\mathrm{AM,2}$ have been adjusted to $\{\tau2^0, \tau_2^1, \kappa_\mathrm{AM,2}\} = \{76, 388, 0.0016\}$. The adjustments are based on the sensitivity analysis shown in Figure~\ref{fig:pe_bar_graph}.}
    \label{fig:pd_comparison}
    \vspace{-0.5cm}
\end{figure}

\subsection{Stochastic Simulation Algorithm}

In this section, the \ac{SSA} solver will be utilized to obtain more nuanced results about the ChemSICal model. For every configuration of parameters, $N_\mathrm{traj} = 500$ trajectories per input concentration were generated to calculate estimates of the metrics described in Section~\ref{subsec:metrics}.
Figure~\ref{fig:pd_comparison} depicts a plot of the input \ac{PMF} and $P_d[\mathbb{N}(Y_\mathrm{on})]$ for the baseline parameter choices, see Table~\ref{tab:initia_values}, in red. We can observe that $P_\mathrm{d}$ varies between 0.5 and 1, with dips towards 0.5 corresponding to the position of the detection thresholds, thereby forming an inverse shape with respect to the input \ac{PMF}. This is the desired behavior, as errors occur primarily for unlikely input values, resulting in an input-weighted error probability of $P_\mathrm{e} = 0.054$ for the baseline ChemSICal configuration. 
This implies that the algorithm generally performs successfully in its chemical form. However, we immediately see that $P_\mathrm{e}\gg P_\mathrm{e,ideal}$, i.e., the ChemSICal model adds significantly more errors to the system compared to the inherent \ac{BEP}. As a consequence, any optimization of the error probability of the ChemSICal model is several orders of magnitude more impactful, even at the expense of increasing the non-chemical inherent error probability $P_\mathrm{e,ideal}$. Therefore, we will look into this in the following.

\subsection{Parameter Adaptation}

\begin{figure}
    \centering
    \includegraphics[width=\linewidth]{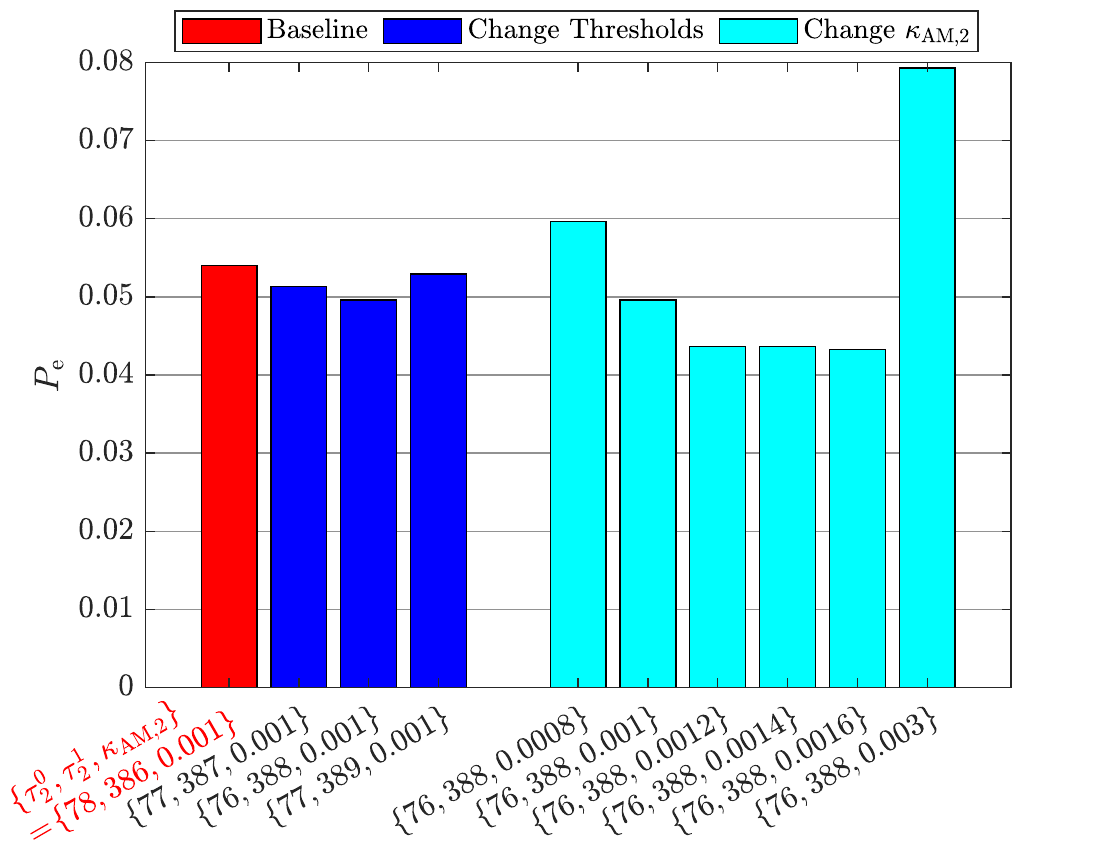}
    \vspace{-0.3cm}
    \caption{Sensitivity analysis of thresholds $\tau_1^0$ and $\tau_1^1$, and \ac{RRC} $\kappa_\mathrm{AM,2}$, which controls the binary decision for TX$_2$. The $y$-axis depicts the input-weighted error probability $P_\mathrm{e}$ for 500 trajectories of the \ac{SSA} solver.}
    \label{fig:pe_bar_graph}
    \vspace{-0.5cm}
\end{figure}

Looking at the baseline $P_\mathrm{d}$ plot in Figure~\ref{fig:pd_comparison}, the most error-prone regions of the input-space are situated around the thresholds, indicating some potential for optimization there. The figure also shows the values of the targeted baseline thresholds, as presented in Table~\ref{tab:parameters}. While the minima of $P_\mathrm{d}$ largely align with the thresholds, they do not match exactly. In fact, a closer look at Figure~\ref{fig:pd_comparison} and an investigation of the trajectories around the thresholds reveals that the \textit{de facto} thresholds, i.e., the values at which the ChemSICal model exhibits a switch in behavior, are slightly higher than desired for $\tau_2^0$, and slightly lower for $\tau_2^1$.
Therefore, we conduct a preliminary analysis of the impacts of changing the thresholds from the suggested optimal values. Figure~\ref{fig:pe_bar_graph} depicts the resulting values of $P_\mathrm{e}$ for a decrease in $\tau_2^0$ and simultaneous increase in $\tau_2^1$, corresponding to changes in $\mathbb{N}(W_\mathrm{2,B})$ and $\mathbb{N}(X_{\mathrm{on},i}), \mathbb{N}(X_{\mathrm{off},i})$. We observe a local optimum configuration with $\{\tau_2^0, \tau_2^1\} = \{76, 388\}$, for which the error probability improves by almost 10\% to 0.049. Figure~\ref{fig:pd_comparison} depicts the resulting $P_d[\mathbb{N}(Y_\mathrm{on})]$ (denoted as \textit{Improved Parameters 1}), where a visible increase of the detection probability around $\tau_1^0$ and in the error-prone region $300<\mathbb{N}(Y_\mathrm{on})<400$ can be observed.

Additionally, we have seen in Section~\ref{subsec:rate_optimization} that the approximate majority reaction block for TX$_2$, and in particular its \ac{RRC} $\kappa_\mathrm{AM,2}$, is  decisive for the performance of the ChemSICal model. Consequently, we will also take a look at a more granular sensitivity analysis around the baseline value. In the following, we will carry over the improved values for the detection thresholds from the prior analysis.
Figure~\ref{fig:pe_bar_graph} shows the resulting $P_\mathrm{e}$ for different values of $\kappa_\mathrm{AM,2}$ in cyan color. We can see that the error probability can be reduced further significantly, by about 10\% to 0.043, if the reaction speed is increased up to $\kappa_\mathrm{AM,2} = 0.0016$. Reducing or further increasing $\kappa_\mathrm{AM,2}$ will drastically worsen the performance of the ChemSICal model, once again underlining the strong effect of even relatively small changes in \acp{RRC} on the performance. Figure~\ref{fig:pd_comparison} also depicts the corresponding $P_\mathrm{d}[\mathbb{N}(Y_\mathrm{on})]$ (denoted as \textit{Improved Parameters 2}). Here, we once again see the largest positive effect in the error-prone range $300<\mathbb{N}(Y_\mathrm{on})<400$. The increased \ac{RRC} leads to a narrower dip in $P_\mathrm{d}$. This only works up to a certain point, as we have seen in Section~\ref{subsec:rate_optimization}, since the approximate majority reaction has to be slow enough for the previous reactions to always deliver the correct results in time.

\section{Conclusion}

This work has presented ChemSICal, a detailed \ac{CRN} model capable of executing the \ac{SIC} algorithm for \ac{MA} within a \ac{DBMC} networking scenario. The \ac{CRN} structure is based on multiple smaller reaction blocks and can run fully simultaneously to extract the transmitted bits from 2 \acp{TX} at the \ac{RX}. Subsequently, both deterministic \ac{ODE} methods and a \ac{SSA} were used to investigate the performance of ChemSICal, revealing the behavior of the system under various input conditions. The \acp{RRC} were identified as crucial parameters to ensure to correct execution of the algorithm and comparing several iteratively adapted values, a reasonable set of \acp{RRC} was found. As the error probability introduced by ChemSICal is found to be much larger than the underlying analytical one, the potential for optimizing the chemical system even at the cost of worsening the significantly smaller inherent \ac{BEP} was underlined. A preliminary sensitivity analysis of detection thresholds and \acp{RRC} showed that the chemical implementation of an algorithm does not necessarily work optimally using the analytical optima found for the non-chemical implementation, but that we need to reevaluate our choices with the chemical behavior of the system in mind, leading to a co-optimization approach between the analytical and the chemical model. 

Further work must focus on generalizing and scaling the presented concepts. Approaches such as timing mechanisms and chemical clocks~\cite{heinleinClosingImplementationGap2024b} could enable faster and more controlled execution of the \ac{CRN}. This could open up opportunities for connecting multiple \acp{CRN}, such that, for example, the parameter optimization of the thresholds could also be implemented chemically. Additionally, the repeated execution and, therefore, the reset of \acp{CRN} must also be taken into account going forward.

\bibliographystyle{ieeetr}
\bibliography{references}

\end{document}